\documentclass[aps,prb,twocolumn,showpacs]{revtex4}

\usepackage{amsmath}
\usepackage{color}
\usepackage{amssymb}

\usepackage{graphicx}

\begin{document}

\title{Comment on "Anisotropic Critical Magnetic Fluctuations in the Ferromagnetic Superconductor UCoGe"}

\author{V.P.Mineev and V.P.Michal}
\affiliation{Service de Physique Statistique, Magn\'{e}tisme et
Supraconductivit\'{e}, Institut Nanosciences et Cryog\'{e}nie, UMR-E CEA/UJF-Grenoble1, F-38054 Grenoble, France}

\begin{abstract}

\end{abstract}
\pacs{75.40.Gb, 74.70.Tx, 75.50.Cc}

\date{\today}
\maketitle
%\section{Introduction}
The results of neutron scattering measurements of magnetic fluctuations in weakly ferromagnetic superconductor UCoGe have been reported  in a recent Letter \cite{Stock}.  There was observed finite attenuation of excitations at zero wave vector earlier found also in another related ferromagnet UGe$_2$ \cite{Huxley} that  has been interpreted by the authors as {\it "strong non-Landau damping of excitations"}. Here we point out  that revealed phenomenon can be treated as {\it the Landau damping} corresponding to the intersection of Fermi surfaces relating to different bands.

The intensity of neutron scattering is proportional to the imaginary part of susceptibility (see for instance \cite{Huxley}).   In the isotropic case the susceptibility is scalar and its imaginary part is given by
\begin{equation}
\frac{ \chi^{\prime\prime}({\bf q},\omega)}{\omega}=\chi({\bf q})\frac{\Gamma_{\bf q}}{\omega^2+\Gamma_{\bf q}^2},
\end{equation}
\begin{equation}
\chi({\bf q})\propto\frac{\chi_pk_F^2}{\xi^{-2}+q^2},
\label{chi}
\end{equation}
Here, $\chi_p$ is the Pauli susceptibility and the line width  is determined by equality
\begin{equation}
\Gamma_{\bf q}\chi({\bf q})=\chi_p\omega({\bf q}) 
\label{gam}
\end{equation}
where $\omega({\bf q})$  is the Landau damping frequency.

Taking into account the possibility of  band intersection one can define the Landau damping frequency  through the imaginary part of bubble diagram with one electron Green's functions
\begin{equation}
\frac{N_0\omega}{\omega_{\nu\nu^\prime}({\bf q})}\propto
\mbox{Im}~T\sum_{{\bf k},\omega_n} G_\nu({\bf k},\omega_n)G_{\nu{^\prime}}({\bf k}+{\bf q},\omega_n+\nu_m)|_{i\nu_m\to\omega+i0},
\end{equation}
where $\nu$ and $\nu^{\prime}$ are the band indices and $N_0$ is average density of states at the Fermi surface. 
For the intraband case $\nu=\nu^{\prime}$  the Landau damping frequency
$\omega({\bf q})\approx v_Fq$
vanishes linearly at $q\to 0$. For the case when the Fermi surfaces of two bands intersect each other  along a line $l$
the Landau damping  at $q=0$ acquires finite value \cite{Smith}
\begin{equation}
\frac{N_0}{\omega_{12}({\bf q}=0)}\approx{\oint\frac{dl}{(2\pi)^3|{\bf v}_1\times{\bf v}_2|}}.
\label{gamS}
\end{equation}
Here vectors ${\bf v}_1$ and ${\bf v}_2$ are the Fermi velocities on the Fermi sheets 1 and 2  at point ${\bf k}_l$ at line $l$.

At ${\bf q}=0$ one can estimate the product (\ref{gam}) as follows $\Gamma_{\bf q}\chi({\bf q})|_{{\bf q}=0}=\chi_p\omega_{12}({\bf q})|_{{\bf q}=0}\approx\chi_p\varepsilon_F$. Numerically this value is of the order of  $10^{-2} K$ that is in correspondence with the experimentally found  values 
of $0.7~\mu$eV and $0.4~\mu$eV  in UGe$_2$ \cite{Huxley} and UCoGe
\cite{Stock} correspondingly.  

The intersection of the different band Fermi surfaces can be established from the ab initio calculations. The latter have been already performed\cite{Biasini,Divis,Mora,Samsel} although without special attention to this problem.

We have presented the potential explanation of nonvanishing at $q=0$ Landau damping measured experimentally in ferromagnetic compounds
UGe$_2$ \cite{Huxley} and UCoGe
\cite{Stock}  based on possible intersection of the Fermi sheets corresponding different bands.  
Quite large and nonvanishing  at $q=0$ value of  the Landau damping  means that the amplitude of pairing interaction  is determined by frequency independent susceptibility.  The latter of  course  should not be taken as it is in the isotropic case given by  equation (\ref{chi}).

\end{document}